\documentstyle[aps,prl,multicol,epsf]{revtex}

\begin{document}

\title{Discrimination between the superconducting gap and the pseudo-gap
in Bi2212 from intrinsic tunneling spectroscopy in magnetic field}

\author{ V.M.Krasnov$^{1,2}$, A.E.Kovalev$^2$, A.Yurgens$^{1,3}$ and D.Winkler$^{1,4}$}

\address{$^1$ Department of Microelectronics and Nanoscience,
 Chalmers University of Technology, S-41296 G\"oteborg, Sweden}

\address{$^2$ Institute of Solid State Physics, 142432 Chernogolovka,
Russia}

\address{$^3$ P.L.Kapitza Institute, 117334 Moscow, Russia}

\address{$^4$ IMEGO Institute, Aschebergsgatan 46, S41133, G\"oteborg,
Sweden}

\date{\today }
\maketitle

\begin{abstract}

Intrinsic tunneling spectroscopy in high magnetic field ($H$) is
used for a direct test of superconducting features in a
quasiparticle density of states of high-$T_c$ superconductors. We
were able to distinguish with a great clarity two co-existing
gaps: (i) the superconducting gap, which closes as $H \rightarrow
H_{c2}(T)$ and $T\rightarrow T_c(H)$, and (ii) the $c$-axis
pseudo-gap, which does not change neither with $H$, nor $T$.
Strikingly different magnetic field dependencies, together with
previously observed different temperature dependencies of the two
gaps ~\cite{Krasnov}, speak against the superconducting origin of
the pseudo-gap.


\end{abstract}

\begin{multicols}{2}

A pseudo-gap (PG) in the electronic density of states (DOS) of
high-$T_c$ superconductors (HTSC) has been established by
different experimental
techniques~\cite{Krasnov,Loeser,Puchkov,Renner,Mihail,Loram,Carreta,Mitrovic,Gorny,Zheng},
at temperatures well above the superconducting critical
temperature $T_c$. Surface tunneling
measurements~\cite{Loeser,Renner} indicated that the
superconducting gap (SG) is almost temperature ($T$) independent,
merging into the pseudo-gap at $T\approx T_c$; the latter can
exist up to room temperature. In contrast to ordinary
superconductors, the gap in DOS of HTSC was reported to be
uncorrelation to $T_c$  and increased with underdoping despite a
decrease in $T_c$~\cite{Miyak}. This has lead to a suggestion that
electron pairs are preformed at high temperature but do not
condense into a coherent state until at $T_c$~\cite{Rand}. Surface
tunneling, however, has drawbacks in its sensitivity to surface
deterioration~\cite{Mallet}. At present there is no consensus
about the origin of the PG, the correlation between SG and PG, or
the dependencies of both gaps on material and experimental
parameters. A clarification of these issues is certainly important
for understanding HTSC.

The most crucial test for the superconducting origin of the two
gaps is their magnetic field ($H$) dependencies. Magnetic field is
a strong depairing factor and destroys superconductivity when the
field exceeds the upper critical field $H_{c2}$. So far the
magnetic field dependence of the pseudo-gap in HTSC is highly
controversial. For example, in YBaCuO compounds the NMR
spin-lattice relaxation rate above $T_c$, which reflects the spin
part of the PG, was reported to decrease~\cite{Carreta},
increase~\cite{Mitrovic}, or to be independent of magnetic
field~\cite{Gorny,Zheng}. Surface tunneling measurements in Bi2212
revealed the existence of the PG inside the vortex
core~\cite{Renner2}. However, the authors argued that it is due to
the precursor superconductivity mechanism, similar, as they
believe, to what happens at $T>T_c$. Clearly, the magnetic field
dependence of the pseudo-gap is far from being established and
advanced measurements with alternative technique are necessary.

Intrinsic tunneling spectroscopy is a powerful method to study the
quasiparticle DOS inside bulk HTSC single crystals
~\cite{Krasnov,Schlenga,Suzuki}. This method allows avoiding the
problem of surface deterioration. On the other hand, it has
several problems too, such as stacking faults (defects) in the
sample, injection of non-equilibrium excitations or internal
heating~\cite{Gough}. Using small area mesa structures, it is
possible to avoid mixing between out-of-plane (c-axis) and
in-plane transport. Moreover, a probability of having stacking
faults in the mesa decreases, as well as the overheating. Thus
clean and clear c-axis tunneling characteristics can be obtained
and used for studying DOS~\cite{Krasnov}.

In this paper we study magnetic field dependencies of the
superconducting gap and the pseudo-gap in small
Bi$_2$Sr$_2$CaCu$_2$O$_{8+x}$ (Bi2212) mesas. The intrinsic
tunneling characteristics exhibit two distinct features: a sharp
superconducting peak and a smooth dip-and-hump feature, attributed
to the $c$-axis pseudo-gap in the tunneling DOS. We have observed
that the SG and the PG not only may have different magnitudes, but
also have strikingly different dependencies both versus $H$ and
$T$: the SG closes as $H \rightarrow H_{c2}$ and $T \rightarrow
T_c$; while the PG does not change neither with $H$, nor $T$ and
persists in the superconducting state. This unambiguously
indicates the coexistence of the two gaps and rules out the
precursor superconductivity scenario of the PG.

Small mesa structures, of area down to 2~${\rm \mu m^2}$ were made
on top of Bi2212 single crystals by photo-lithography, Ar-ion
etching and a self-alignment technique~\cite{Krasnov}. Such mesas
represent stacks of intrinsic Josephson junction, formed by the
crystalline structure of HTSC~\cite{Muller} and exhibit behavior
typical for stacks of SIS
(Superconductor-Insulator-Superconductor) type Josephson
junctions. For example, Fiske steps and Fraunhofer oscillations
were observed in Ref.~\cite{Fiske} and multiple Josephson fluxon
modes were observed in

\begin{figure}
\noindent
\begin{minipage}{0.48\textwidth}
\epsfxsize=0.9\hsize \centerline{ \epsfbox{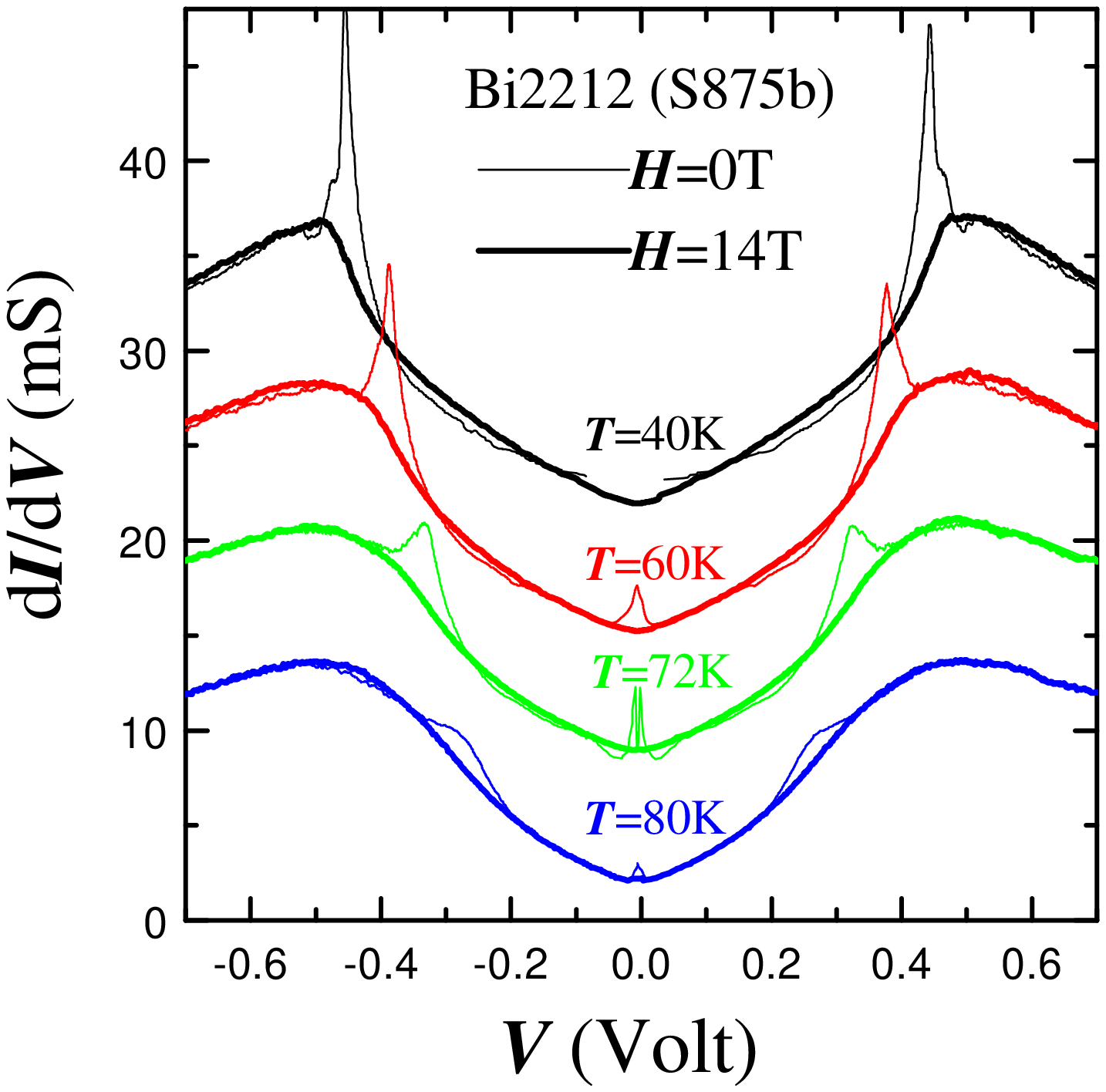} }
\vspace*{6pt} \caption{Differential conductance curves for a pure
Bi2212 mesa at different temperatures for $H=$0 (thin lines) and
$H=14T$ (thick lines). The existence of a $(T,H)$ dependent
superconducting peak and the $(T,H)$ independent "background"
pseudo-gap dip-and-hump is clearly seen}
\end{minipage}
\end{figure}

\noindent Ref.~\cite{Modes}. Here we present results for pure
Bi2212 and intercalated HgBr$_2$-Bi2212 samples. The Bi2212
samples are near optimally doped ($T_c \simeq$ 92 K). The
intercalated samples are overdoped, as follows from a small PG and
more metallic temperature dependence of the $c$-axis resistance.
Upon intercalation with HgBr$_2$ molecules, the distance between
cuprate layers increases and the $c$-axis resistivity increases
while the critical current density decreases by a factor of 10-20
as compared to the pristine sample~\cite{Inter}. This leads to a
decrease of both dissipated power and overheating during
measurements. $T_c$ is decreased by intercalation to $\sim$73~K.
The $c$-axis IVC's were measured in a three probe configuration.
The differential conductance, $\sigma=dI/dV$, was measured by
lock-in technique with the AC current of 10$\mu$A.

In Fig.~1, $c$-axis $\sigma = dI/dV$ vs. $V$, curves are shown for
a pure Bi2212 mesa ($5\times 6.5 \mu m^2$, $N=7$ intrinsic
Josephson junctions) at four different temperatures, $T<T_c$.
Curves are displayed both for $H$=0 (thin lines) and $H$=14T
(thick lines) along the $c$-axis. For clarity, the curves for
different $T$ are shifted sequentially by 7 mS along the vertical
axis.

The $\sigma (V)$ curves have a temperature independent resistance
at large bias~\cite{Krasnov}, as expected for pure tunnel
junctions. At low $T$ and $H$ there is a sharp superconducting
peak at the sum-gap voltage, $V_s=2N\Delta_S/e$, where $N$ is the
number of junctions in the mesa and $\Delta_s$ is the
superconducting gap~\cite{dwave&H}. For SIS junctions, the
superconducting peak should disappear simultaneously with
superconductivity both as $T \rightarrow T_c$ and $H \rightarrow
H_{c2}$~\cite{Golubov}. Indeed, from Fig. 1 it is seen that with
increasing $T$, the peak shifts to smaller voltages, reduces in
amplitude and

\begin{figure}
\noindent
\begin{minipage}{0.48\textwidth}
\epsfxsize=0.9\hsize \centerline{ \epsfbox{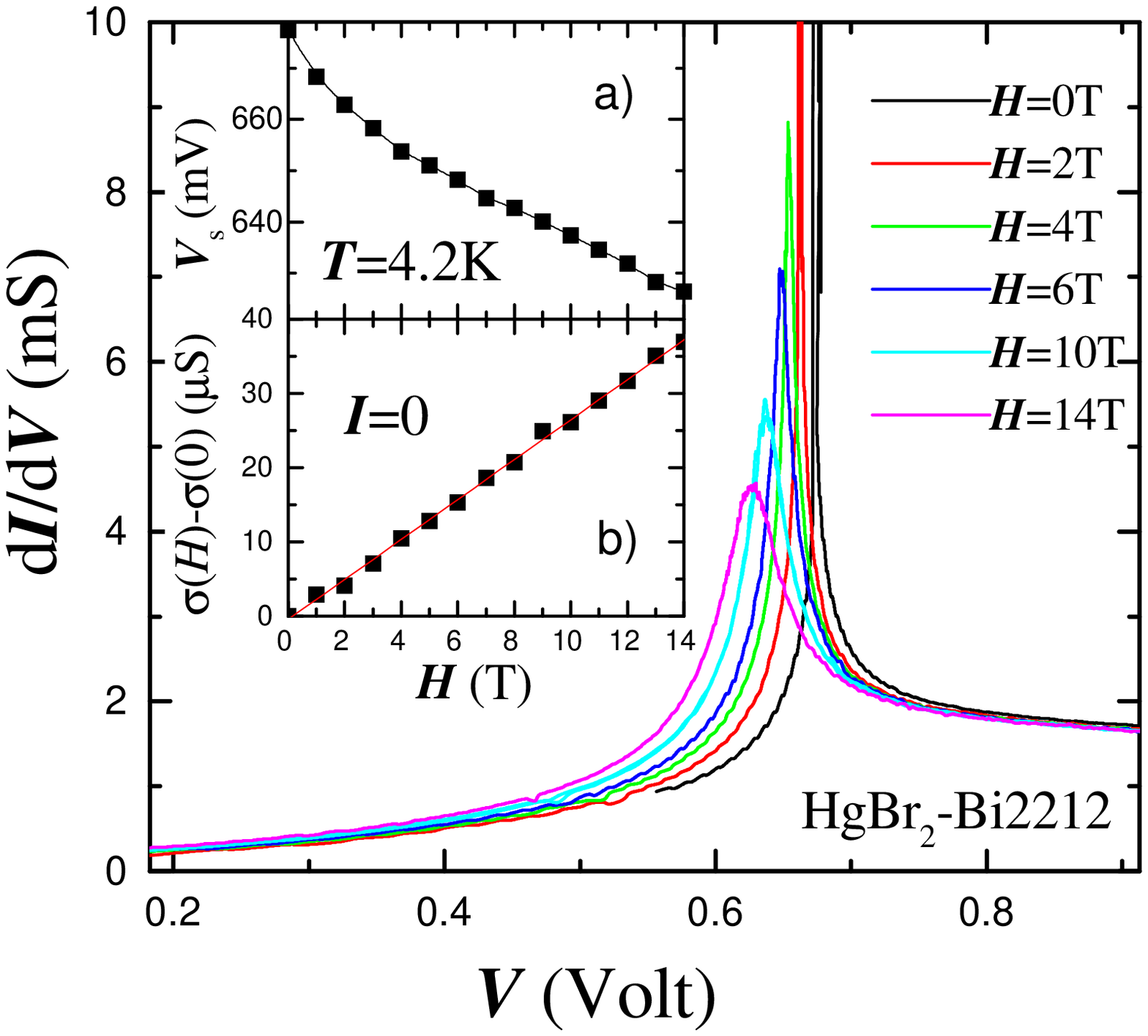} }
\vspace*{6pt} \caption{Differential conductance of the
intercalated mesa at $T=$4.2K, for different $H
\parallel c$. Insets show a) the decrease of the peak
voltage, $V_s$, and b) the linear increase of a zero-bias
conductance with $H$.}
\end{minipage}
\end{figure}

\noindent eventually disappears as $T_c$ is
approached~\cite{Krasnov}. It is also seen that the
superconducting peak is very sensitive to magnetic field and
vanishes at large fields.

In Fig. 2, $\sigma (V)$ curves of the intercalated HgBr$_2$-Bi2212
mesa are shown for different $H \parallel c$ at low temperature
$T$=4.2K. It is seen that the superconducting peak reduces in
amplitude with increasing field and systematically shifts to a
lower voltage, as shown in inset a). Likewise, the zero bias
conductance, $\sigma (I=0)$, increases linearly with $H$, showing
negative magnetoresistance, see inset b). Such behavior is typical
for SIS junctions at low temperature in perpendicular magnetic
fields both for $s$-wave~\cite{Golubov} and $d$-wave
superconductors with uncorrelated vortices~\cite{Morozov}. Here,
the decrease of $V_s$ reflects the decrease of the maximal
$\Delta_s$ and the linear increase of $\sigma (I=0)$ is associated
with a linear increase of the vortex density.

In Fig. 3, $\sigma (V)$ curves of the same pure Bi2212 mesa as in
Fig. 1. are shown for high temperatures, a) $T\simeq$72K and b)
$T\simeq$80K, and for several $H\parallel c$. Unlike the case of
low $T$, for $T>70$K the superconducting peak is completely
suppressed at $H=$14 T. With increasing $H$, the $\sigma (V)$
curves approach some magnetic field independent "background"
curve, e.g., from Fig. 3 b) it is clearly seen that there are no
changes in the $\sigma (V)$ curves for $H>$10T. Taking the field
at which the superconducting peak disappears and the
magnetoresistance at $V \sim V_s$ saturates as the upper critical
field, we estimate $H_{c2}(\parallel c) \simeq$ 10T at $T=$80 K
and $\simeq$ 14T at $T=$72K, in agreement with previously obtained
values, see, e.g., Ref.~\cite{Morozov}.

However, despite vanishing of the superconducting peak, the IVC's
remain nonlinear at $H>H_{c2}(T)$. The background $\sigma (V)$
curves exhibit a smooth depletion of the zero-bias conductance (a
dip) plus a hump at a certain voltage $V=V_{pg}$, see Fig. 1. The
background curve

\begin{figure}
\noindent
\begin{minipage}{0.48\textwidth}
\epsfxsize=0.9\hsize \centerline{ \epsfbox{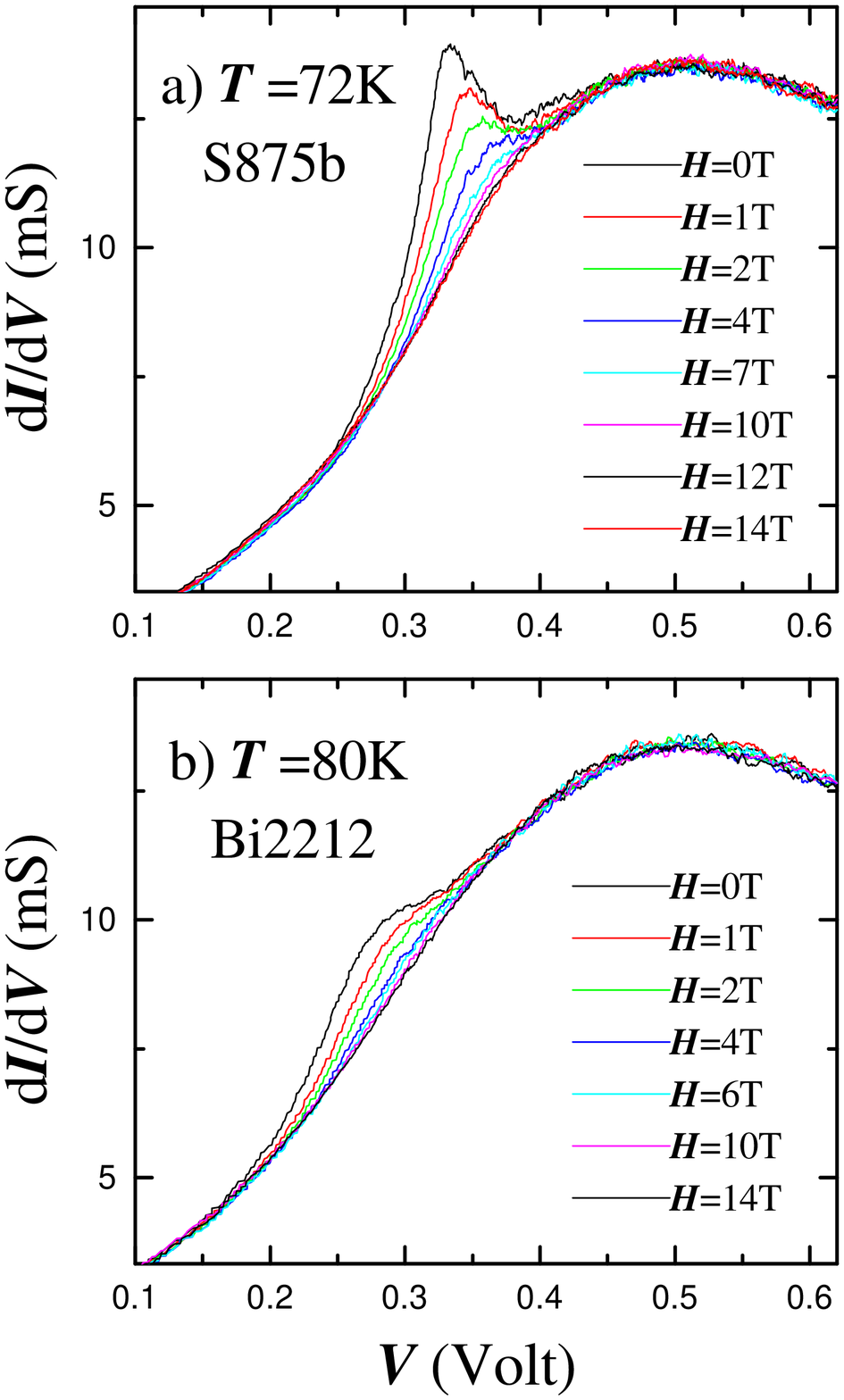} }
\vspace*{6pt} \caption{$\sigma(V)$ curves of a pure Bi2212 mesa at
a) $T=$72K and b) $T=$ 80 K and for several $H
\parallel c$. It is seen that the
superconducting peak is completely suppressed in high fields,
while the "background" PG dip-and-hump structure remains magnetic
field independent. It is clear that the superconducting peak and
the PG hump appear at different voltages.}
\end{minipage}
\end{figure}

\noindent is similar to that observed in the vortex core
\cite{Renner2} and, to our opinion, represents the bare normal
(non-superconducting) state with the remaining PG in the tunneling
DOS even at $T<T_c$. Fig. 1 demonstrates that the PG hump voltage
$V_{pg}$ is almost independent of $T$~\cite{Krasnov} and is
completely independent of $H$ within the applied field range. The
latter is most clearly demonstrated in Fig. 3 b), from which it is
seen that there are no changes neither of the shape nor of the
position of the PG hump with $H$. Moreover, it is seen that the PG
hump voltage, $V_{pg}$, is distinctly different from the
superconducting peak voltage, $V_s$. The striking contrast in
magnetic field dependencies of the superconducting gap and the
pseudo-gap is the central observation of this paper.

At high $T$, the behavior of $\sigma (V)$ becomes more
complicated: First, the superconducting peak does not move to
lower voltages with increasing $H$ and second, the subgap
conductance, $\sigma (0)$, becomes almost independent of

\begin{figure}
\noindent
\begin{minipage}{0.48\textwidth}
\epsfxsize=0.9\hsize \centerline{ \epsfbox{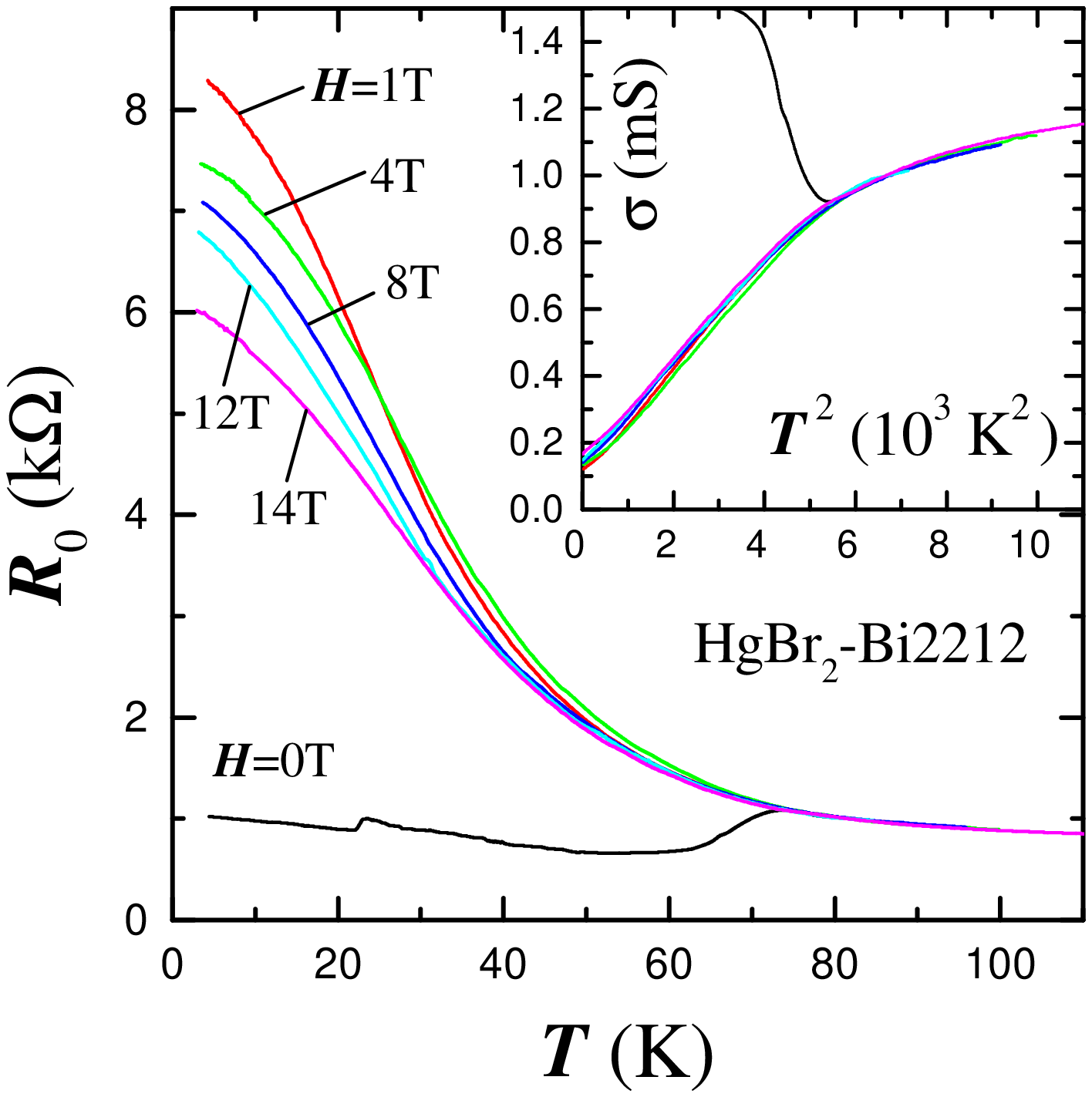} }
\vspace*{6pt} \caption{Temperature dependence of the zero-bias
resistance of the intercalated mesa at several $H
\parallel c$. It is seen that $R_0$ decreases with increasing $H$
and approaches a universal magnetic field independent "background"
curve. Inset demonstrates the $T^2$ dependence of the zero-bias
conductance at low temperatures.}
\end{minipage}
\end{figure}

\noindent $H$. To some extent such anomalous behavior may be due
to the fact that at high $T$ and $H$, the $\sigma(V)$ of a SIS
junction does not follow the quasi-particle DOS~\cite{dwave&H}. As
a matter of fact, for $T/T_c \gtrsim 0.5$ the peak in $\sigma (V)$
can shift to a higher voltage, as a result of a strong smearing,
despite the decrease of $\Delta _s$~\cite{Golubov}. In addition,
the peak in the spatially averaged DOS itself smears and shifts
towards higher energies with $H$, see, e.g., Fig. 18 from
Ref.\cite{Ichioka}. Similarly, the $\sigma (0)$ can even decrease
with $H$ despite the increase of DOS$(E=0)$~\cite{Golubov}.
However, numerical simulations showed that the subgap conductance
of a SIS junction should still have some $H$-dependence, which is
obviously not the case for $V<$ 0.25V in Fig.3 a).

We believe that the observed discrepancy between numerical
simulations and experimental data at high $T$ and $H$, is due to
the PG, persisting at $T<T_c$. Indeed, the phenomenon appears at
high $T$, when the $T$-dependent SG becomes considerably less than
the $T$-independent PG. Furthermore, the anomalous behaviour is
more pronounced for the near optimally doped Bi2212 with a larger
PG, than for overdoped HgBr$_2$-Bi2212 samples with a smaller PG.
If so, the suppression of subgap conductance at $T>$ 60K, for the
pure Bi2212 mesa, see Fig. 1, is mostly due to the PG rather than
to the SG.

Fig. 4, shows temperature dependencies of the zero-bias resistance
$R_0$ of the intercalated sample at several $H$. The intercalated
sample is remarkable in the respect, that the critical current can
be completely suppressed by a small $H$. Thus it was possible to
trace the quasiparticle resistance in the whole range of $T$ and
$H$. At low $T$, the $R_0$ continuously decreases ($\sigma(0)$
increases) with field, as shown in inset b) of Fig. 2. The range
of variation of $R_0(H)$ decreases with increasing $T$. At high
$H$ and $T$, e.g. in Fig.4 for $H>$12 T and $T>$35 K, the $R_0(T)$
curves collapse into a single semiconducting curve. This
"background", magnetic field independent resistance represents the
persisting PG, rather than the SG, as discussed above. The inset
in Fig. 4 shows the zero-bias conductance vs. $T$- square for the
same data. It is seen that $\sigma(0)\propto T^2$, at low $T$. In
Ref.\cite{Morozov} such a behavior was attributed to a $d$-wave
character of the superconducting state. However, we should
emphasize that in our case the zero-bias conductance is not
entirely determined by the SG, but a significant part of it
originates from the magnetic field independent "background" PG
conductance. The inset in Fig. 4 shows that the background
conductance at $H=$14T and $T>$30K is still reasonably well
described by the $T^2$ dependence. This may indicate that the PG
also has a $d$-wave symmetry \cite{Loeser}. The behavior of
$R_0(T,H)$ of pure, near optimally doped Bi2212 mesas is
similar~\cite{Yurgens}, but quantitatively different. First of all
the PG in those samples is significantly larger than in overdoped
intercalated HgBr$_2$-Bi2212 samples, in agreement with a general
trend, known from previous
studies~\cite{Krasnov,Loeser,Puchkov,Renner,Mihail,Loram}.
Therefore, the magnetic field independent "background" PG
resistance is a larger fraction of $R_0$ and the magnetoresistance
at zero bias decreases considerably, as can be seen from Figs. 1
and 3. We should also note, that for optimally doped mesas
$\sigma(0)$ deviates from the $T^2$ dependence and shows a
tendency to vanish at $T\rightarrow 0$.

In conclusion, we have performed intrinsic tunneling spectroscopy
of small HTSC mesa structures in high magnetic fields. Such
measurements yield a crucial test for the high-$T_c$
superconductivity, since a high magnetic field is a strong
depairing factor. The intrinsic tunneling conductance curves
exhibit two distinct features: (i) the sharp peak at low
temperatures and (ii) the smooth dip-and-hump structure at high
temperatures. As we have shown, the two features can be clearly
separated due to their strikingly different magnetic field
dependencies: the low-$T$ peak vanishes both as $H \rightarrow
H_{c2}(T)$ and $T\rightarrow T_c(H)$, while the dip-and-hump does
not change with neither $H$, nor $T$. This allowed us to
unambiguously attribute the peak to the superconducting gap and
the dip-and-hump to the non-superconducting $c$-axis pseudo-gap in
tunneling DOS. Our data indicate that the PG neither disappear,
nor transform into the SG below $T_c$. Indeed, at intermediate $T$
and $H$, the superconducting peak emerges on top of the PG
dip-and-hump feature at a distinctly lower voltage, $V_s<V_{pg}$,
see Fig. 3. Furthermore, at high $H$ we were able to completely
suppress the superconducting peak/gap and observed the bare
pseudo-gap state clearly persisting at $T$ well below $T_c(H=0)$.
In experiment, such state is characterized by saturation of the
magnetoresistance and appearance of the "background",
$(T,H)$-independent nonlinear conductance, see Figs. 3 and 4. The
obtained results speak in favor of two different co-existing gaps
in the tunneling DOS of HTSC. This casts serious doubts to the
precursor superconductivity scenarios of the $c$-axis pseudo-gap
and related HTSC theories, which require a similar origin and
behaviour of the SG and the PG. We hope that our observation
together with the growing experimental evidence for an independent
behavior of SG and
PG~\cite{Krasnov,Puchkov,Mihail,Loram,Gorny,Zheng}, will help
consolidating efforts to understand the pseudo-gap phenomenon and
the high-$T_c$ superconductivity.

The work was supported by the Swedish Superconductivity
Consortium and NFR. SnL (G\"oteborg) facilities were used for sample
fabrication. Magnetic field measurements were performed at ISSP
(Chernogolovka). We are grateful to V.V.Ryazanov for the
support of this work, to A.A.Golubov for the assistance in
numerical simulations, to J.-H.~Choy for providing us with intercalated crystals,
to T.Claeson and P.Delsing for critical reading the manuscript.

\end{multicols}

\begin{references}
\bibitem{dwave&H} We measure spatially averaged characteristics. In the mixed state,
$\Delta_s$ is spatially nonuniform due to a presence of Abricosov
vortices. At low $T$ the peak represents the maximal
superconducting gap, $\Delta_s(max)$, between the vortices.
However, at higher $T$ the correlation between the peak position
and $\Delta_s(max)$ is lost and $V_s$ becomes systematically
larger than $2\Delta_s(max)/e$.
\end{references}
\end{document}